\pdfoutput = 1
\documentclass[
 aps,
 prd,
 reprint,
 nofootinbib,
 nobibnotes,
 superscriptaddress,
 preprintnumbers,
 groupedaddress
]{revtex4-2}

\usepackage[revtex]{hep-paper}
\graphics{figures}

\usepackage{isotope}
\usepackage{dcolumn}

\acronym{SM}{Standard Model}
\acronym{LLP}{long-lived particle}
\acronym{DM}{dark matter}
\acronym{QGP}{quark–gluon plasma}
\acronym{LHC}{Large Hadron Collider}
\acronym[HL-LHC]{HLLHC}{high-luminosity phase of the accelerator}
\acronym{PDF}{parton distribution function}
\acronym{EMD}{electromagnetic dissociation}
\acronym{BFPP}{bound-free pair production}
\acronym[$\nu$MSM]{nuMSM}{Neutrino Minimal Standard Model}
\acronym{NLO}{next-to-leading order}
\acronym{NLL}{next-to-leading log}
\acronym{BAU}{baryon asymmetry of the universe}

\begin{document}

\preprint{CP3-19-26}

\title{New Long Lived Particle Searches in Heavy Ion Collisions at the LHC}

\author{Marco Drewes}
\email{marco.drewes@uclouvain.be}
\author{Andrea Giammanco}
\email{andrea.giammanco@uclouvain.be}
\author{Jan Hajer}
\email{jan.hajer@uclouvain.be}
\author{Michele Lucente}
\email{michele.lucente@uclouvain.be}

\affiliation{Centre for Cosmology, Particle Physics and Phenomenology, Université catholique de Louvain, Louvain-la-Neuve B-1348, Belgium}

\begin{abstract}
We show that heavy ion collisions at the LHC provide a promising environment to search for signatures with displaced vertices in well-motivated New Physics scenarios.
Compared to proton collisions, they offer several advantages,
\begin{inlinelist}
\item[\ref{it:interactions}] the number of parton level interactions per collision is larger
\item[\ref{it:pileup}] there is no pile-up
\item[\ref{it:trigger}] the lower instantaneous luminosity compared to proton collisions allows to operate the LHC experiments with very loose triggers
\item[\ref{it:fields}] there are new production mechanisms that are absent in proton collisions
\end{inlinelist}
In the present work we focus on the third point and show that the modification of the triggers alone can increase the number of observable events by orders of magnitude if the long lived particles are predominantly produced with low transverse momentum.
Our results show that collisions of ions lighter than lead are well-motivated from the viewpoint of searches for New Physics.
We illustrate this for the example of heavy neutrinos in the Neutrino Minimal Standard Model.
\end{abstract}

\maketitle

\section{Introduction}

The search for new elementary particles is one of the most pressing quests in contemporary physics.
While the \SM of particle physics describes phenomena ranging from the inner structure of the proton to the spectra of distant galaxies with an astonishing accuracy, there are unambiguous proofs that it does not constitute a complete description of Nature, including the \DM puzzle, the excess of matter over antimatter in the observable universe and flavor-changing oscillations among the known neutrino states.
In addition to these facts, there a number of observations that in principle can be explained by the \SM and the theory of General Relativity, but only by choosing values of the fundamental parameters that are considered unsatisfactory by many theorists, including the \emph{electroweak hierarchy problem} \cite{tHooft:1979rat}, the \emph{flavor puzzle} \cite{Weinberg:1977hb}, \emph{strong CP problem} \cite{Belavin:1975fg, tHooft:1976rip, Jackiw:1976pf} and the value of the vacuum energy or \emph{cosmological constant} \cite{Zeldovich:1967gd, Weinberg:1988cp}.
Before the \LHC was turned on it was widely believed, based on theoretical arguments \cite{Giudice:2008bi}, that the new particles that are responsible for these phenomena have masses only slightly above the electroweak scale and interact with the \SM particles at a rate that allows to produce them copiously at the \LHC.
However, the non-observation of any new elementary particles beyond the Higgs boson predicted by the \SM has broadened the scope of experimental searches at CERN \cite{Beacham:2019nyx} and other laboratories around the world.
In the present situation it seems mandatory, from an experimental viewpoint, to minimize the dependence on specific theory frameworks and \enquote{turn every stone} in the search for New Physics.
It has recently been pointed out that this strategy should also include the use of data from heavy ion collisions to search for new phenomena \cite{Bruce:2018yzs}.

In the present work we study the possibility to search for \LLPs in heavy ion collisions.
\LLPs appear in a wide range of models of physics beyond the \SM \cite{Alimena:2019zri}.
They can owe their longevity to a combination of different mechanisms that are known to be at work in the \SM, including symmetries, the mass spectrum, and small coupling constants.
In collider experiments they reveal themselves with \emph{displaced signatures} in the detectors.
The displacement makes it possible to distinguish the signal from the many tracks environment that is created in a heavy ion collision, because all primary tracks from primary \SM interactions originate from within the microscopic volume of the two colliding nuclei.
Our proposal is driven by the approach to fully exploit the discovery potential of the existing detectors to make optimal use of CERN's resources and infrastructure.
This approach is complementary to proposals that aim to improve the sensitivity of the \LHC to \LLPs by adding new detectors, including the recently approved FASER experiment \cite{Feng:2017uoz} and other dedicated detectors, such as MATHUSLA \cite{Chou:2016lxi, Curtin:2018mvb, Alpigiani:2018fgd}, CODEX-b \cite{Gligorov:2017nwh} and Al3X \cite{Gligorov:2018vkc}.
A summary of our main results is presented in reference \cite{Drewes:2018xma}.

\section{Heavy ion collisions at the LHC} \label{sec:heavy ion}

For an equal number of collisions and equal center-of-mass energy per nucleon, collisions of heavier nuclei guarantee larger hard-scattering cross sections than proton ($pp$) collisions, thanks to the enhancement factor of $\approx A^2$ in the number of parton level interactions, where $A$ is the mass number of the isotope under consideration.
In the case of lead isotopes (\isotope[208][82]{Pb}) accelerated in the \LHC, $A = 208$ provides an enhancement of four orders of magnitude.
However, there are several drawbacks.
\begin{enumerate}[label = \arabic*)]

\begin{figure}
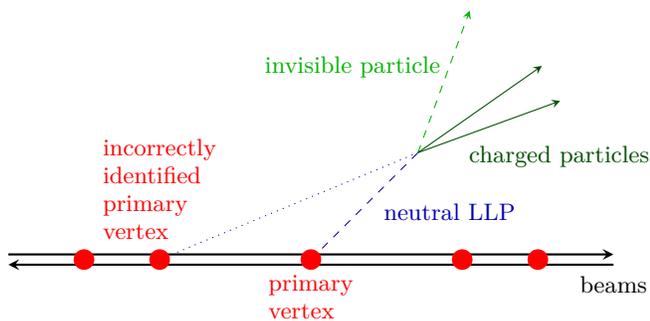

\graphic{misidentification}
\caption{
Example of a signature that is difficult to search for in high pileup $pp$ collisions.
Heavy ion collisions can provide a cleaner environment.
} \label{fig:misidentification}
\end{figure}

\begin{table}
\renewcommand{\arraystretch}{1.15}
\newcommand{\cent}[1]{\multicolumn{1}{c}{#1}}
\begin{tabular}{r@{}l*{8}{r@{ }}}
 \toprule
& & \cent{$M_\text{ion}$} & \cent{$\sqrt{s_{XX}}$} & \cent{$\sigma_{\EMD}$} & \cent{$\sigma_{\BFPP}$} & \cent{$\sigma_\text{had}$} & \cent{$\sigma_\text{tot}$} & \cent{$\sigma^\text{XX}_W$} & \cent{$\sigma_W$} \\
& & \cent{[GeV]} & \cent{[TeV]} & \cent{[b]} & \cent{[b]} & \cent{[b]} & \cent{[b]} & \cent{[nb]} & \cent{[\mu b]} \\
\midrule
\isotope[1][1]{} & H & 0.931 & 14.0 & 0 & 0 & 0.071 & 0.071 & 20.3 & 0.0203 \\
\isotope[16][8]{} & O & 14.9 & 7.00 & 0.074 & $24 \cdot 10^{-6}$ & 1.41 & 1.47 & 10.0 & 2.56 \\
\isotope[40][18]{} & Ar & 37.3 & 6.30 & 1.2 & 0.0069 & 2.6 & 3.81 & 8.92 & 14.3 \\
\isotope[40][20]{} & Ca & 37.3 & 7.00 & 1.6 & 0.014 & 2.6 & 4.21 & 10.0 & 16.0 \\
\isotope[78][36]{} & Kr & 72.7 & 6.46 & 12 & 0.88 & 4.06 & 17.0 & 9.16 & 55.7 \\
\isotope[84][36]{} & Kr & 78.2 & 6.00 & 13 & 0.88 & 4.26 & 18.2 & 8.43 & 59.5 \\
\isotope[129][54]{} & Xe & 120 & 5.86 & 52 & 15 & 5.67 & 72.7 & 8.22 & 137 \\
\isotope[208][82]{} & Pb & 194 & 5.52 & 220 & 280 & 7.8 & 508 & 7.69 & 333 \\
\bottomrule
\end{tabular}
\caption{
\resetacronym{EMD}
\resetacronym{BFPP}
Cross sections for different heavy ions based on \cite{Citron:2018lsq}.
Here $M_\text{ion}$ indicates the mass of the ion and $\sqrt{s_\text{XX}}$ the nucleon-nucleon center of mass energy achievable at the \LHC.
The total cross section is the sum of the \EMD, the \BFPP and the hadronic cross section $\sigma_\text{tot} = \sigma_{\EMD}  + \sigma_{\BFPP} + \sigma_\text{had}$.
$\sigma^\text{XX}_W$ indicates the $W$ boson cross section per nucleon.
For this illustrative purpose of this table we estimate $\sigma^\text{XX}_W$ from $p p \to W^\pm \to \mu^\pm \nu$ with $p_T(\mu) > \unit[5]{GeV}$, calculated at \NLO using \software{MadGraph5\_aMC@NLO} \cite{Alwall:2014hca}, while we properly simulate heavy ion collisions as described below to obtain the results shown in \cref{fig:W boson sensitivity}.
$\sigma_W$ indicates the nuclear cross section, estimated by scaling $\sigma^\text{XX}_W$ with $A^2$.
} \label{tab:cross sections}
\end{table}

\item The collision energy per nucleon in heavy ion collisions is smaller than in proton collisions.
The design \LHC beam energy is limited to \unit[2.76]{TeV} for Pb ions (corresponding to a center-of-mass energy per nucleon of $\sqrt{s_\text{XX}} = \unit[5.52]{TeV}$ in PbPb collisions), to be compared with \unit[7]{TeV} for proton beams (hence $\sqrt{s_{pp}} = \unit[14]{TeV}$).
These design values are expected to be reached during Run~3.
The scaling factor $\flatfrac{\eval*{\sigma^{pp}}_{\unit[14]{TeV}}}{\eval*{\sigma^\text{XX}}_{\unit[5.52]{TeV}}}$ grows as a function of the particle masses in the final state of the hard process under consideration \cite{Cartiglia:2013vsa}, and it is typically larger for gluon-initiated processes than for quark-antiquark collisions.
For instance, for top quark pair ($t\bar t$) production the drop between \unit[14]{TeV} and \unit[5.52]{TeV} is of one order of magnitude \cite{Sirunyan:2017ule} due to the large mass of the top quark.
However, the $W$-boson production cross section per nucleon is only reduced by a factor of around 2.5 (\cf \cref{tab:cross sections}).
The reduction factor is around 2.3 for $B$ mesons \cite{Cacciari:1998it, Cacciari:2001td, Cacciari:2015fta}, which are lighter but whose production at these \LHC energies is mostly initiated by gluon-gluon fusion (opposed to $W$ bosons, which are mostly created by quark-antiquark annihilation).

\item Heavy ion collisions are characterized by the production of a very large particle multiplicity, and in particular a very large multiplicity of charged-particle tracks, which poses challenges for data acquisition and data analysis to the multipurpose \LHC experiments.
While a large track multiplicity generates a strong background for prompt signatures, the decay of feebly interacting \LLPs produces displaced tracks at macroscopic distances from the interaction point that can be easily distinguished from the tracks that originate from the primary interaction at the collision point.
We address this point in \cref{sec:multiplicities}.

\begin{table*}
\renewcommand{\arraystretch}{1.15}
\newcolumntype{.}[1]{D{.}{.}{#1}}
\newcommand{\cent}[1]{\multicolumn{1}{c}{#1}}
\newcommand{\symbols}{\cent{$\mathscr L_0$} & \cent{$\tau_b$} & \cent{$\mathscr L_\text{ave}$} & \cent{$\nicefrac{N_\text{XX}}{N_{pp}}$}}
\newcommand{\units}{\cent{[\inv{\mu bs}]} & \cent{[h]} & \cent{[\inv{\mu bs}]} & \cent{[1]}}
\begin{tabular}{r@{}l@{ }*{3}{.{6}@{ }.{3}@{}.{8}@{ }l}@{}}
 \toprule
 & & \multicolumn{4}{c}{pessimistic ($p = 1$)} & \multicolumn{4}{c}{realistic ($p = 1.5$)}& \multicolumn{4}{c}{optimistic ($p = 1.9$)}
\\ \cmidrule(r){3-6} \cmidrule(lr){7-10} \cmidrule(l){11-14}
 & & \symbols & \symbols & \symbols
\\ & & \units & \units & \units
\\ \midrule
 \isotope[1][1]{}& H & 21.0 \cdot 10^3 & 75.0 & 15.0 \cdot 10^3 & 1
 & 21.0 \cdot 10^3 & 75.0 & 15.0 \cdot 10^3 & 1
 & 21.0 \cdot 10^3 & 75.0 & 15.0 \cdot 10^3 & 1
\\
 \isotope[16][8]{} & O & 1.43 & 52.6 & 1.07 & 0.0082
 & 14.6 & 16.4 & 8.97 & 0.0688
& 94.3 & 6.48 & 45.5 & 0.349
\\
 \isotope[40][18]{} & Ar & 0.282 & 45.8 & 0.208 & 0.00889
& 1.29 & 21.5 & 0.837 & 0.0358
& 4.33 & 11.7 & 2.46 & 0.105
\\
 \isotope[40][20]{} & Ca & 0.229 & 46.0 & 0.168 & 0.00811
 & 0.937 & 22.7 & 0.615 & 0.0296
 & 2.90 & 12.9 & 1.69 & 0.0811
\\
 \isotope[78][36]{} & Kr & 0.0706 & 20.6 & 0.0454 & 0.00758
& 0.161 & 13.6 & 0.0948 & 0.0158
 & 0.311 & 9.80 & 0.169 & 0.0282
\\
 \isotope[84][36]{} & Kr & 0.0706 & 19.2 & 0.0448 & 0.00797
& 0.161 & 12.7 & 0.0933 & 0.0166
& 0.311 & 9.15 & 0.166 & 0.0296
\\
 \isotope[129][54]{} & Xe & 0.0314 & 7.20 & 0.156 & 0.00637
 & 0.0476 & 5.84 & 0.0222 & 0.00908
 & 0.0665 & 4.94 & 0.0294 & 0.0120
\\
 \isotope[208][82]{} & Pb & 0.0136 & 1.57 & 3.79 \cdot 10^{-3} & 0.00379
& 0.0136 & 1.57 & 3.8 \cdot 10^{-3} & 0.00379
& 0.0136 & 1.57 & 3.8 \cdot 10^{-3} & 0.00379
\\
\bottomrule
\end{tabular}
\caption{
Luminosities for collisions of different heavy ions based on \cite{Citron:2018lsq} for three choices of the scaling parameter $p$ (\cf definition \eqref{eq:scaling law}).
$\mathscr L_0$ is the peak luminosity, $\tau_b$ the optimal beam lifetime, and $\mathscr L_\text{ave}$ the optimized average luminosity.
The last column contains the ratio between the number of events $N_\text{XX} = L \sigma_W$ in XX- and $pp$-production, where $L$ is the integrated luminosity (\cf definition \eqref{eq:integrated luminosity}) and $\sigma_W$ is given in \cref{tab:cross sections}.
Following \cite{Citron:2018lsq}, we use an optimistic turnaround time of \unit[1.25]{h}, which we compensate in the case of heavy ion collisions by assuming that the useful run time is only half of the complete run time.
} \label{tab:luminosities}
\end{table*}

\item The instantaneous luminosity in heavy ion runs is limited to considerably lower values compared to $pp$ collisions, \cf \cref{tab:luminosities}.
The \LHC delivered \unit[1.8]{\inv{nb}} of collisions to the ATLAS and CMS detectors during the latest PbPb Run in late 2018, and \unit[10]{\inv{nb}} are expected to be accumulated during the \HLLHC \cite{Citron:2018lsq}.
In terms of sheer numbers, this cannot compete with the size of the $pp$ data samples even if the $A^2$ enhancement due to nucleon-nucleon combinatorics is taken into account.
This poses the strongest limitation when searching for rare phenomena in heavy ion data.
We discuss these luminosity limitations in some detail in \cref{sec:luminosity}.

\item Heavy ion Runs at the \LHC are comparably short.
In the past not more than one month has been allocated in the yearly schedule, as opposed to around six months in the $pp$ case.
This is not a fundamental restriction, and one can imagine that the sharing of time may change in the future depending on the priorities of the \LHC experiments.
In the following we compare the sensitivity per equal running time, given a realistic instantaneous luminosity, in order to remain independent of possible changes in the planning.
\end{enumerate}
On the other hand, there are key advantages in heavy ion collisions.
\begin{enumerate}[label = \roman*)]

\item \label{it:interactions} The aforementioned number of parton level interactions per collision is larger.

\item \label{it:pileup} There is no pileup in heavy ion collisions.
In collisions of high intensity proton beams pileup leads to tracks that originate from different points in the same bunch crossing and creates a considerable background for displaced signatures.
In heavy ion collisions the probability of mis-identifying the primary vertex is negligible.
Hence, heavy ion collisions can provide a cleaner environment to search for signatures stemming from the decay of \LLPs when pileup is a problem, \cf \cref{fig:misidentification}.

\item \label{it:trigger} The lower instantaneous luminosity makes it possible to operate ATLAS and CMS with significantly lower trigger thresholds.
This, \eg, allows to detect events with comparably low transverse momentum $p_T$ in scenarios involving light mediators or when the \LLPs are produced in the decay of mesons.

\item \label{it:fields}
In heavy ion collisions there are entirely new production mechanisms are absent or inefficient in proton collisions.
The strong electromagnetic fields in heavy-ion collisions can drastically increase the production cross section for some exotic states that couple to photons.
This can be exploited by considering ultraperipheral heavy ion collisions \cite{Baltz:2007kq}, as emphasized in recent publications on monopoles \cite{Gould:2017zwi} and axion like particles \cite{Knapen:2016moh}.
It has also been suggested that thermal processes in the \QGP can help to produce a sizable number of exotic states \cite{Bruce:2018yzs}.
\end{enumerate}
This article presents an illustrative study with an analysis strategy based entirely on \cref{it:interactions,it:trigger}.
The effect of \cref{it:pileup} is model dependent, and explained in more detail in \cref{sec:multiplicities}.
A detailed quantitative analysis of the effects deriving from \cref{it:pileup} goes beyond the scope of the present article, whose main purpose is to point out the potential of heavy ion collisions for \LLP searches.
We do not explore the (strongly model dependent) \cref{it:fields} in the present work.
A list of references on this topic can \eg be found in reference \cite{Bruce:2018yzs}.

\section{Track and vertex multiplicities} \label{sec:multiplicities}

Historically, heavy ion collisions have been considered an overly complicated environment, therefore unsuitable for precise measurements of particle properties or searches of rare phenomena, because of their large final-state particle multiplicity, as opposed to $pp$ collisions.
However, due to the high pileup during Run~4 in $pp$ collisions, the track multiplicity is expected to become comparable for $pp$ and PbPb collisions while it is smaller for lighter ion collisions \cite{Sirunyan:2019cgy}.

In PbPb collisions, hard-scattering signals are more likely to originate in the most central events, where up to around 2\,000 charged particles are produced per unit of rapidity at $\sqrt{s_\text{XX}} = \unit[5.52]{TeV}$ \cite{Adam:2015ptt}, meaning that around 10\,000 tracks can be found in the tracking acceptance of the multi-purpose experiments ATLAS and CMS.
In contrast, $pp$ collisions during standard Runs in 2017 were typically overlaid by about 30 pileup events, each adding about 25 charged particles on average within the tracking acceptance of the multi-purpose detectors \cite{Khachatryan:2015jna, Aaboud:2016itf, Adam:2015pza}, meaning that $\approx 750$ charged particles per event are coming from pileup.
This is not expected to increase by a large factor until the end of Run~3.
The \HLLHC will bring a big jump: current projections assume that, in order to accumulate \unit[3\,000]{\inv{fb}} as planned, each bunch crossing will be accompanied by about 200 pileup events \cite{Apollinari:2015bam, Apollinari:2017cqg}, meaning 5\,000 additional charged particles per hard-scattering event.
In conclusion, in the \HLLHC era the difference in track multiplicity between most-central PbPb and $pp$ collisions will reduce to a mere factor of two.
A lot of ingenuity has been invested by the major LHC experiments in recent years to overcome the issues deriving from such a large track multiplicity \cite{CERN-LHCC-2015-020, CMSCollaboration:2015zni}.
In addition to the planned detector upgrades, all particle reconstruction and identification algorithms have been made more robust and optimized for a regime of very large multiplicities, and these efforts automatically benefit also the analysis of heavy ion data.

Although a very large track multiplicity is expected to degrade the reconstruction and identification of displaced vertices, the adverse effect of pileup on vertex-finding performance is caused more by the presence of additional primary-interaction vertices than from the sheer number of tracks.
This is \eg demonstrated by the comparison of $b$-tagging performance in $t\bar t$ studies in $pp$, $p$Pb, and PbPb collisions \cite{SiRunyan:2017xku, CMS-PAS-HIN-19-001}.
Using the same algorithm as in the standard $pp$ analysis, and requiring an equal efficiency of correctly tagged $b$-quark-initiated jets, the misidentification rate of light jets is smaller in $p$Pb than in $pp$ events (\unit[0.1]{\%} \vs \unit[0.8]{\%}) in spite of the larger track multiplicity \cite{SiRunyan:2017xku}.
However, in the case of PbPb collisions a dedicated retuning of the $b$-tagging algorithm was necessary in order to recover a comparable efficiency.
Nevertheless an acceptable purity versus efficiency (sufficient to provide evidence for top quark production) was achieved even for the most central collisions, in which the track multiplicity is maximal \cite{CMS-PAS-HIN-19-001}.
Similar qualitative considerations apply to the case of algorithms for the reconstruction of long-lived particles.

\section{Average instantaneous luminosity} \label{sec:luminosity}

The maximum luminosity achievable in heavy ion collisions is constrained by multiple factors.
\begin{enumerate}[label = \arabic*)]

\item Technical limits set on the injector performance.

\resetacronym{EMD}
\resetacronym{BFPP}

\item The total cross section per nucleon is increased compared to $pp$ collisions due to the additional sizable electromagnetic contributions.
This results in a more rapid decline of the beam intensity.
Moreover, most of the interactions are unwanted electromagnetic interactions caused by the stronger electromagnetic fields and soft hadronic processes, \ie, \EMD and \BFPP, \cf \eg references \cite{Schaumann:2015mvg, Braun:2014naa, Jowett:2015dmf} and references therein for details.
The change of the mass/charge ratio caused by these processes leads to secondary beams that can potentially quench the LHC magnets.
This problem was only recently mitigated for ATLAS and CMS by directing the secondary beams between magnets, while a special new collimator is required for ALICE \cite{Jowett:1977371, Jowett:2018yqk}.

\item Collecting the maximum rate of events that the LHC can deliver is not necessarily ideal for all the experiments.
For instance, the ALICE experiment is limited in the amount of data that it can acquire by the repetition time of its time projection chamber \cite{Aamodt:2008zz}, thus instantaneous luminosity is leveled at their interaction point by adjusting the horizontal separation between the bunches.
Similarly also the LHCb experiment only uses about \unit[10]{\%} of the available beam intensity \cite{Zhang:2016hmo}.

\end{enumerate}
The upper limit on the achievable instantaneous luminosity depends on the charge $Z$ and mass $A$ of the accelerated nuclei in a complicated manner and is currently under investigation.
For the purpose of the present article, we use the numbers presented in \cref{tab:luminosities}, which are computed based on estimates presented at a recent \HLLHC workshop \cite{Jowett:2018jj}, \cf also \cite{Citron:2018lsq}.
In the following, we briefly summarize how we used these data.
The instantaneous luminosity at one interaction point (IP) scales according to \cite{Benedikt:2015mpa}
\begin{equation}
\mathscr L \propto n_b N_b^2
\ ,\label{eq:luminosity}
\end{equation}
where $n_b$ is the number of bunches per beam and $N_b$ is the number of nucleons per bunch.
The decay of the beam due to interactions follows
\begin{equation}
 \dv{N_b}{t}
 = - \sigma_\text{tot} \frac{n_\text{IP}}{n_b} \mathscr L
 = - \frac{N_b^2}{N_0 \tau_b}
\ ,
\end{equation}
where $n_\text{IP}$ is the number of interaction points, $\sigma_\text{tot}$ is the total cross section, $N_0 = N_b(0)$ is the initial number per bunch at beam injection and
\begin{equation}
\tau_b
 = \frac{n_b}{\sigma_\text{tot} n_\text{IP}} \frac{N_0}{\mathscr L_0}
\propto \frac{1}{\sigma_\text{tot} n_\text{IP} N_0}
\ ,\label{eq:lifetime}
\end{equation}
is the beam lifetime.
Here $\mathscr L_0$ is the initial instantaneous luminosity at beam injection.
Therefore, the number of nucleons per bunch decays according to
\begin{align}
 N_b(t)
 & = \frac{N_0}{1 + \theta}
\ ,& \text{with}&
 & \theta
 & = \frac{t}{\tau_b}
\ .\label{eq:evolution}
\end{align}
The evolution of the instantaneous luminosity $\mathscr L(t)$ and integrated luminosity $L(t)$ are then
\begin{align}
 \mathscr L(t)
 & = \frac{\mathscr L_0}{\left(1 + \theta\right)^2}
\ ,&
 L(t)
 & = \mathscr L_0 \tau_b \frac{\theta}{1 + \theta}
\ .\label{eq:integrated luminosity}
\end{align}
The turnaround time $t_\text{ta}$ is the average time between two physics runs.
Therefore, the average luminosity is
\begin{equation}
 \mathscr L_\text{ave}(t) = \frac{L(t)}{t + t_\text{ta}}
\ ,
\end{equation}
which is maximized for
\begin{align}
 t_\text{opt}
 & = \tau_b \sqrt{\theta_\text{ta}}
 \ ,
 & \text{with}&
 & \theta_\text{ta}
 & = \frac{t_\text{ta}}{\tau_b}
 \ .
\end{align}
Finally, the average luminosity for the optimal run time is
\begin{equation}
 \mathscr L_\text{ave}(t_\text{opt})
 = \frac{\mathscr L_0}{\left(1 + \sqrt{\theta_\text{ta}}\right)^2}
 \ .
\end{equation}
Additionally, the initial bunch intensity follows roughly
\begin{equation}
N_b \left(\isotope[A][Z]{X}\right) = N_b \left(\isotope[208][82]{Pb}\right) \left(\frac{Z}{82}\right)^{-p}
\ ,\label{eq:scaling law}
\end{equation}
where the exponent characterizes the number of nucleons per bunch.
For a given isotope, it is limited by the heavy-ion injector chain, the bunch charges and intra-beam scatterings.
Simple estimates based on fixed target studies with Ar beams suggest that $1 \lesssim p \lesssim 1.9$ is realistic \cite{Jowett:2018jj}.

\section{An example: Heavy Neutrinos}

\begin{figure}
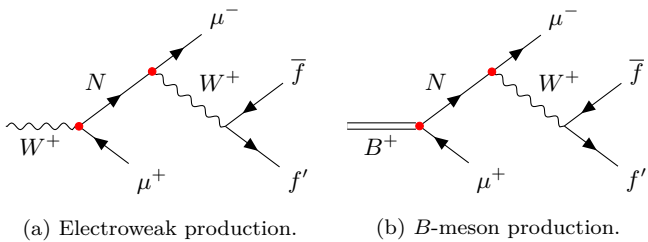

\begin{subfigure}{.48\linewidth}
\graphic{W-Production}
\caption{
Electroweak production.
} \label{fig:W boson diagram}
\end{subfigure}
\hfill
\begin{subfigure}{.48\linewidth}
\graphic{B-Production}
\caption{
$B$-meson production.
} \label{fig:B meson diagram}
\end{subfigure}
\caption{
Production and decay of heavy neutrino mass eigenstates $N$.
The tiny couplings inducing the displaced signature are indicated by red vertices.
} \label{fig:feynman diagrams}
\end{figure}

In the following, we use the example of heavy neutrinos with masses below the electroweak scale that interact with the \SM exclusively through their mixing with ordinary neutrinos, to illustrate the potential of New Physics searches in heavy ion collisions.
This is an extremely conservative approach for two reasons.
First, we do not take advantage of any of the new production mechanisms that the strong electromagnetic fields or the \QGP offer in comparison to proton collisions, \cf \cref{it:fields}.
Second, we do not take advantage of the lack of pileup, \cref{it:pileup}, which we do not expect to play a major role in the minimal seesaw model considered here.
This point can, however, give heavy ion collisions a crucial advantage over proton runs in searches for signatures with a more complicated topology than the decays shown in \cref{fig:feynman diagrams}.
In the context of heavy neutrinos this could \eg be the case in left-right symmetric models \cite{Pati:1974yy, Mohapatra:1974gc, Senjanovic:1975rk} where decays mediated by Majorons can lead to pairs of displaced vertices \cite{Nemevsek:2016enw}.

Right handed neutrinos $\nu_R$ appear in many extensions of the \SM.
The implications of their existence strongly depend on the values of their masses.
The $\nu_R$ could solve several open puzzles in cosmology and particle physics, \cf \eg \cite{Drewes:2013gca} for a review.
Most notably they can explain the light neutrino masses via the type-I seesaw mechanism \cite{Minkowski:1977sc, GellMann:1980vs, Mohapatra:1979ia, Yanagida:1980xy, Schechter:1980gr, Schechter:1981cv}, which requires one flavor of $\nu_R$ for each non-zero neutrino mass in the \SM.
In addition they may explain the \BAU via leptogenesis \cite{Fukugita:1986hr}, act as \DM candidates \cite{Dodelson:1993je}, address various anomalies observed in neutrino oscillation experiments \cite{Abazajian:2012ys} or generate the Higgs potential radiatively \cite{Brivio:2017dfq}.
\footnote{
For further details we refer the reader to the following reviews on the matter-antimatter asymmetry \cite{Canetti:2012zc}, the perspectives to test leptogenesis \cite{Chun:2017spz}, sterile neutrino \DM \cite{Adhikari:2016bei, Boyarsky:2018tvu} and experimental searches for heavy neutrinos \cite{Atre:2009rg, Deppisch:2015qwa, Cai:2017mow, Antusch:2016ejd}.
\label{fn:reviews} }

Since right handed neutrinos are gauge singlets, the number of their flavors is not constrained to be equal to the \SM generations by anomaly considerations.
At least two flavors of $\nu_R$ are required to explain the observed neutrino oscillation data via the seesaw mechanism.
Here we work in a simple toy model with only a single flavor of $\nu_R$ with mass $M$, which is sufficient because the displaced vertex signature does not rely on interference effects among different neutrinos or correlations between their parameters.
The minimal extension of the \SM with right handed neutrinos can be obtained by adding all renormalizable operators that only contain $\nu_R$ and \SM fields to the \SM Lagrangian,
\begin{equation}
 \mathcal L_{\nu_R} =
 \frac{\operatorname i}{2} \overline{\nu_R} \slashed \partial \nu_R
- F_a \overline{\ell_L}_a \varepsilon \phi^* \nu_R
- \frac{1}{2} \overline{\nu_R^c} M \nu_R
+ \text{h.c.}
\ .\label{eq:Lagrangian}
\end{equation}
Here $\phi$ is the \SM Higgs doublet, $\ell_{La}$ are the \SM lepton doublets, the $F_a$ are the Yukawa coupling constants to the \SM lepton generation $a$ and $\varepsilon$ is the antisymmetric SU(2) tensor.

The heavy neutrino interactions with the \SM can be described by the mixing angles $\theta_a = \flatfrac{v F_a}{M}$ with $v = \ev{\phi} \simeq \unit[174]{GeV}$, which characterize the relative suppression of their weak interactions compared to those of the light neutrinos.
The Lagrangian that describes the interaction of the heavy neutrino mass eigenstate $N \simeq \nu_R + \theta_a \nu_{La}^c + \text{c.c.}$ with the \SM reads
\begin{multline}
 \mathcal L
 \supset
- \frac{m_W}{v} \overline N \theta^*_a \gamma^\mu e_{L a} W^+_\mu
- \frac{m_Z}{\sqrt 2 v} \overline N \theta^*_a \gamma^\mu \nu_{L a} Z_\mu \\
- \frac{M}{v} \theta_a h \overline{\nu_L}_\alpha N
+ \text{h.c.}
\ ,\label{eq:weak intraction}
\end{multline}
where $h$ is the physical Higgs field after spontaneous breaking of the electroweak symmetry.
The mixing angles $U_a^2 \equiv \abs{\theta_a}^2$ can be large enough to produce sizable numbers of heavy neutrinos in collider experiments if the heavy neutrinos approximately respect a generalized $\text{B} - \text{L}$ symmetry \cite{Gluza:2002vs,Shaposhnikov:2006nn,Kersten:2007vk}, where B and L denote baryon and lepton number, respectively (\cf also \cite{Moffat:2017feq}).

\begin{figure}
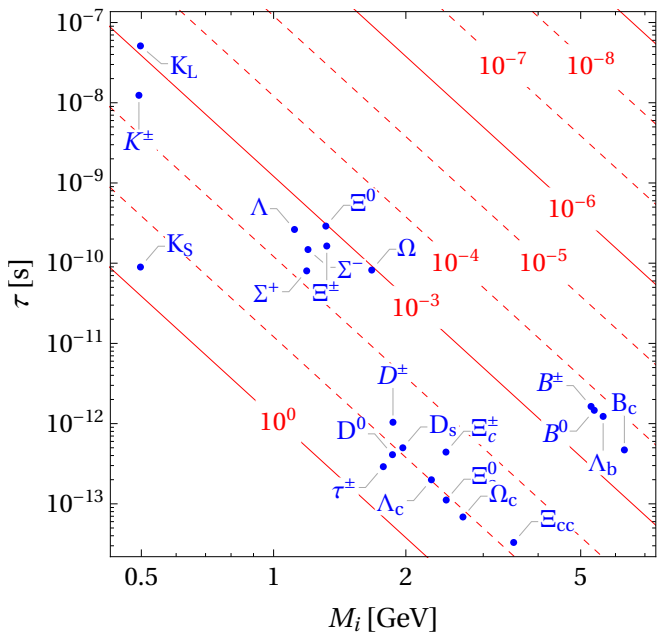

\graphic{background}
\caption{
Heavy neutrino mixing $U^2$ (red lines) compared to potentially relevant \SM backgrounds (blue dots) as a function of particle mass $M_i$ and lifetime $\tau$.
Figure taken from reference \cite{Drewes:2019fou}.
In addition to these \SM backgrounds, secondary nuclear interactions in the detector material constitute a source of background for displaced signatures.
For comparable track multiplicity the effect of the \SM backgrounds would, depending on the particle's lifetime, be equal or worse in $pp$ data compared to heavy ion data.
} \label{fig:background}
\end{figure}

For $M$ below the weak gauge boson masses, the heavy neutrinos can be long-lived enough to produce displaced vertex signals at the LHC \cite{Dib:2014iga, Helo:2013esa, Izaguirre:2015pga, Gago:2015vma, Dib:2015oka, Cvetic:2016fbv, Cottin:2018kmq, Antusch:2017hhu, Cottin:2018nms, Abada:2018sfh, Drewes:2019fou, Drewes:2018xma, Bondarenko:2019tss, Liu:2019ayx, Dib:2019ztn, Dib:2018iyr, Cvetic:2018elt, Cvetic:2019rms, Cvetic:2019shl} or at future collider \cite{Blondel:2014bra, Antusch:2015mia, Antusch:2016ejd, Antusch:2016vyf, Antusch:2017pkq},
\footnote{
Such searches could be much more sensitive in models where the heavy neutrinos have additional interactions \cite{Graesser:2007pc, Graesser:2007yj, Maiezza:2015lza, Batell:2016zod, Nemevsek:2016enw, Caputo:2017pit, Nemevsek:2018bbt, Cottin:2019drg}.
Heavy ion collisions can be a promising place to search for signatures with two displaced vertices, \cf \eg \cite{Nemevsek:2016enw}, that would benefit from the better vertex identification, \cf \cref{it:pileup}.
}
\cf \cref{fig:background}.
In this mass range the Lagrangian \eqref{eq:Lagrangian} effectively describes the phenomenology of the \nuMSM \cite{Asaka:2005an, Asaka:2005pn}, a minimal extension of the \SM that can simultaneously explain the light neutrino masses, \DM and the baryon asymmetry of the universe \cite{Canetti:2012vf, Canetti:2012kh}, \cf \cite{Boyarsky:2009ix} for a review.
The dominant production channel for $M > \unit[5]{GeV}$ is the decay of real $Z$ ($W$) bosons, in which the heavy neutrinos are produced along with a neutrino $\nu_a$ (charged lepton $\ell_a$), while for $M < \unit[5]{GeV}$ the production in $b$-flavored hadron decays dominates, \cf \cref{fig:feynman diagrams}.
The number of heavy neutrinos that are produced along with a lepton of flavor $a$ can be estimated as $\sim L \sigma_\nu U_a^2 $, where $\sigma_\nu$ is the production cross section for light neutrinos.
It is roughly given by $\sigma_\nu \simeq \flatfrac{\sigma_W}{3}$ in $W$-decays and $\sigma_\nu \simeq \flatfrac{\sigma_B}{9}$ in $B$-decays, where $\sigma_W$ and $\sigma_B$ are the $W$ and $B$ production cross sections in a given process.
They then decay semileptonically or purely leptonically, \cf \eg \cite{Gorbunov:2007ak, Atre:2009rg, Canetti:2012kh, Abada:2017jjx, Bondarenko:2018ptm, Pascoli:2018heg}.

\begin{figure}
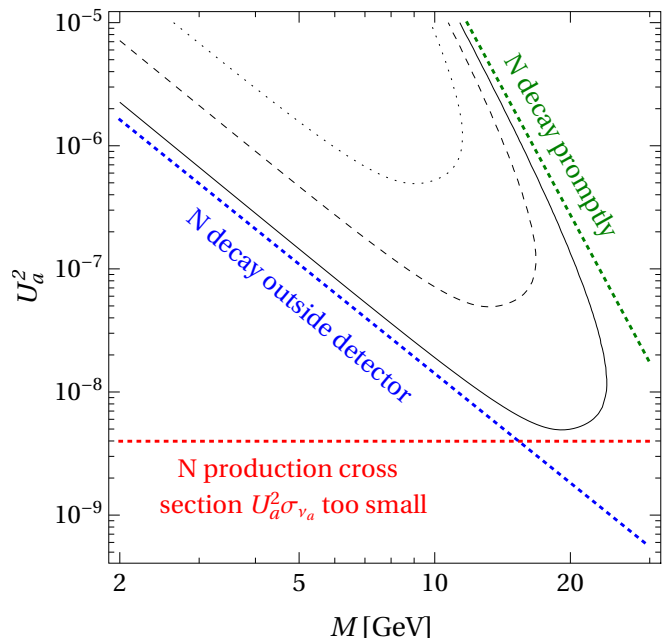

\graphic{illustration}
\caption{
We show a sensitivity estimate for $W$ induced
$N$ decays in $pp$ collisions based on the simplified detector model \eqref{eq:number W events} with $l_0 = \unit[5]{mm}$ and $l_1 = \unit[20]{cm}$.
The three sensitivity curves (black) correspond to nine expected events with integrated luminosities of \unit[3, 30, 300]{\inv{fb}} as a function of heavy neutrino mass $M$ and mixing $U_a^2$.
The colored dotted lines illustrate the three main obstacles in improving the reach.
} \label{fig:illustrative plot}
\end{figure}

The number of displaced vertex events with a lepton of flavor $a$ at the first vertex and a lepton of flavor $b$ from the second vertex that can be seen in a detector can then be estimated by as
\begin{equation}
N_\text{obs} \simeq L \sigma_\nu \frac{U_a^2 U_b^2}{U^2} \left[\exp(-\frac{l_0}{\lambda_N}) - \exp(-\frac{l_1}{\lambda_N}) \right] f_\text{cut}
\ .\label{eq:number W events}
\end{equation}
Here $l_1$ is the length of the effective detector volume in a simplified model of a spherical detector, $l_0$ the minimal displacement that is required by the trigger, $\lambda_N = \flatfrac{\beta \gamma}{\Gamma_N}$ is the particle decay length, where $\Gamma_N$ is the heavy neutrino decay width, $\beta$ is the heavy neutrino velocity and $\beta \gamma = \flatfrac{\abs{\bm p}}{M}$ the usual Lorentz factor, $U^2 = \sum_a U_a^2$ is the total mixing and $f_\text{cut} \in [0, 1]$ is an overall efficiency factor that parameterizes the effects of cuts due to triggers, deviations of the detector geometry from a sphere and detector efficiencies.
The analytic formula \eqref{eq:number W events} allows for an intuitive understanding of the sensitivity curves obtained from simulations, \cf \cref{fig:illustrative plot}.
As illustrated in \cref{fig:errors}, it can reproduce the results of simulated data surprisingly well.

One may wonder whether the heavy neutrinos can leave the dense plasma that surrounds the collision point.
Intuitively, this should clearly be the case because the scattering cross section of heavy neutrinos is suppressed by a factor $\sim U^2$ compared to that of ordinary neutrinos.
For a more quantitative estimate, we can evaluate the mean free path $\lambda_T$ of the relativistic heavy neutrinos of energy $\omega_p \simeq \flatfrac{m_W}{2}$ that are produced in real gauge boson decays as $\lambda_T \simeq \gamma \beta/\Gamma_T$, where $\Gamma_T$ is the thermal damping rate in a plasma of temperature $T$.
In this regime it is known that $\Gamma_T < \flatfrac{U^2 T^2}{\omega_p}$ \cite{Ghiglieri:2016xye}.
We can therefore estimate
\begin{equation}
\thickmuskip=5mu plus 3mu minus 1mu
\lambda_T \simeq \frac{\gamma \beta}{\Gamma_T} > \frac{\abs{\bm p} \omega_p}{U^2 T^2 M} > \frac{\abs{\bm p}^2}{U^2 T^2 M} > \frac{1}{U^2 M} \approx \unit[\frac{\unit[0.2]{GeV}}{U^2M}]{fm}
\ ,
\end{equation}
which is %
orders of magnitude
larger than a few tens of fm.

Since $f_\text{cut}$ is largest for muons, in the following we concentrate on a benchmark model in which the heavy neutrinos mix exclusively with the second generation ($U^2 = U_\mu^2$).
\footnote{
Realistic flavor mixing patterns in the seesaw model in view of current neutrino oscillation data have recently been studied in \cite{Chrzaszcz:2019inj}, we refer the interested reader to this article and references therein.
}
The expression \eqref{eq:number W events} then further reduces to
\begin{equation}
N_\text{obs} \simeq L \sigma_\nu U_\mu^2 \left[\exp(-\frac{l_0}{\lambda_N}) - \exp(-\frac{l_1}{\lambda_N}) \right] f_\text{cut}
\ .
\end{equation}

\subsection{Heavy neutrinos from $W$ boson decay} \label{sec:W boson decays}

\subsubsection{Event generation} \label{sec:simulation}

We first study the perspectives to find heavy neutrinos produced in the decay of $W$ bosons in a displaced vertex search.
Our treatment of the detector closely follows that in reference \cite{Drewes:2019fou}, but we have adapted the simulation of the production for different colliding isotopes.
We calculate the Feynman rules for Lagrangian \eqref{eq:weak intraction} with \software[2.3]{FeynRules} \cite{Alloul:2013bka}, using the implementation \cite{Degrande:2016aje} that is based on the computations in references \cite{Atre:2009rg, Alva:2014gxa}.
Then we generate events for the processes shown in \cref{fig:feynman diagrams} with \software[2.6.4]{MadGraph5\_aMC@NLO} \cite{Alwall:2014hca}, which is capable of generating events for heavy ion collisions if provided with the appropriate \PDFs.
For the simulation of lead collisions we use published \PDFs \cite{Eskola:2016oht}.
However, for Argon and other intermediate ions there are no published \PDFs, therefore
we calculate the ion \PDFs by scaling the proton \PDFs.
The \PDF for a quark of flavor $a$ within a ion with mass number $A = p + n$ is denoted by $f^a_{p, n}(x, Q^2)$, where $x$ is the Bjorken fraction defined as the ratio of the parton energy over the ion energy.
It can be approximated by a re-scaling of the proton \PDF $f_{1, 0}$ via
\begin{subequations}
\begin{align}
f^d_{p, n}(x) &= p f^d_{1, 0}(A x) + n f^u_{1, 0}(A x)\ , \\
f^u_{p, n}(x) &= p f^u_{1, 0}(A x) + n f^d_{1, 0}(A x)\ , \\
f^{\bar d}_{p, n}(x) &= p f^{\bar d}_{1, 0}(A x) + n f^{\bar u}_{1, 0}(A x)\ , \\
f^{\bar u}_{p, n}(x) &= p f^{\bar u}_{1, 0}(A x) + n f^{\bar d}_{1, 0}(A x)\ , \\
f^f_{p, n}(x) &= A f^f_{1, 0}(A x)\ ,
\end{align}
\end{subequations}
where the index $f$ denotes quarks beyond the first generation and gluons.
For the sake of notational clarity we have dropped the scale dependence $Q^2$.

We find that the effects of the nuclear \PDFs can be neglected after comparing them to other sources of uncertainty in our analysis.
We use \software{MadWidth} \cite{Alwall:2014bza} to calculate the $N$ decay width,
\footnote{
\software{MadWidth} uses quarks instead of hadrons in the final states.
The resulting error is relatively small as long as all particles are relativistic, as we explicitly checked by comparison with the results in \cite{Gorbunov:2007ak}.
}
the resulting lifetime is given in \cref{fig:background}.
Subsequently, we simulate the decays with \software{MadSpin} \cite{Frixione:2007zp, Artoisenet:2012st}.
Finally, we hadronize the colored particles and generate hadronic showers with \software[8.2]{Pythia} \cite{Sjostrand:2014zea}.
We calculate the detector efficiencies of the CMS detector using our own code based on public information of the detector geometry.
Most importantly, we use a pseudorapidity coverage of $\abs{\eta} \leq 4$ and use for the extension of the tracker \unit[1.1 and 2.8]{m} in the transversal and longitudinal direction, respectively \cite{CMSCollaboration:2015zni}.
In \cite{Drewes:2019fou} it has been shown that in $pp$ collisions the expected performance of the ATLAS detector is comparable to the one of the CMS detector for this search strategy.
We expect the same to be true in heavy ion collisions.

We search this signal in event samples that have either been triggered by a single muon or by a pair of muons.
The minimal transverse momentum $p_T$ of the muon used for the pair triggers can be softer than in the single muon triggers.
For the tagging and tracking efficiencies we use the CMS detector card values of \software[3.4.1]{DELPHES} \cite{deFavereau:2013fsa}.
In order to reduce the background from long lived \SM hadrons we require that the secondary vertices have a minimal displacement $l_0$ of \unit[5]{mm}.
In order to suppress further backgrounds, in particular from nuclear interactions of hadrons produced in the primary collisions with the detector material, we require at least two displaced tracks with an invariant mass of at least \unit[5]{GeV} in the reconstruction of the displaced vertices, \cf reference \cite{Cottin:2018kmq}.
The reconstruction efficiency is near \unit[100]{\%} if the produced particles traverse the entire tracker.
If a particle traverses only a fraction of the tracker the efficiency is reduced.
We adapt a ray tracing \cite{Smits:1998ei, Williams:2005ae} method to compute the particle's trajectory and use the length of the remaining path within the tracking system as the criterion to estimate the vertex reconstruction efficiency.
It has recently been shown in reference \cite{Aaboud:2017iio} that the detection efficiency drops only linearly with the displacement if advanced algorithms are used.
We adopt this functional dependence and assume that the maximal displacement that can still be detected can be improved by a factor 2 if optimized algorithms are used.
This strategy closely follows the approach taken in reference \cite{Drewes:2019fou}.

We fix the integrated luminosity in PbPb runs to \unit[5]{\inv{nb}}, a realistic value for one month in the heavy ion program.
We then use the relations presented in \cref{sec:luminosity} to estimate the integrated luminosity that could be achieved with Ar in the same period as \unit[0.5 and 5]{\inv{pb}} for pessimistic and optimistic assumptions for the scaling behavior, respectively.
For protons we use \unit[50]{\inv{fb}}.

\subsubsection{Backgrounds}

\begin{figure*}
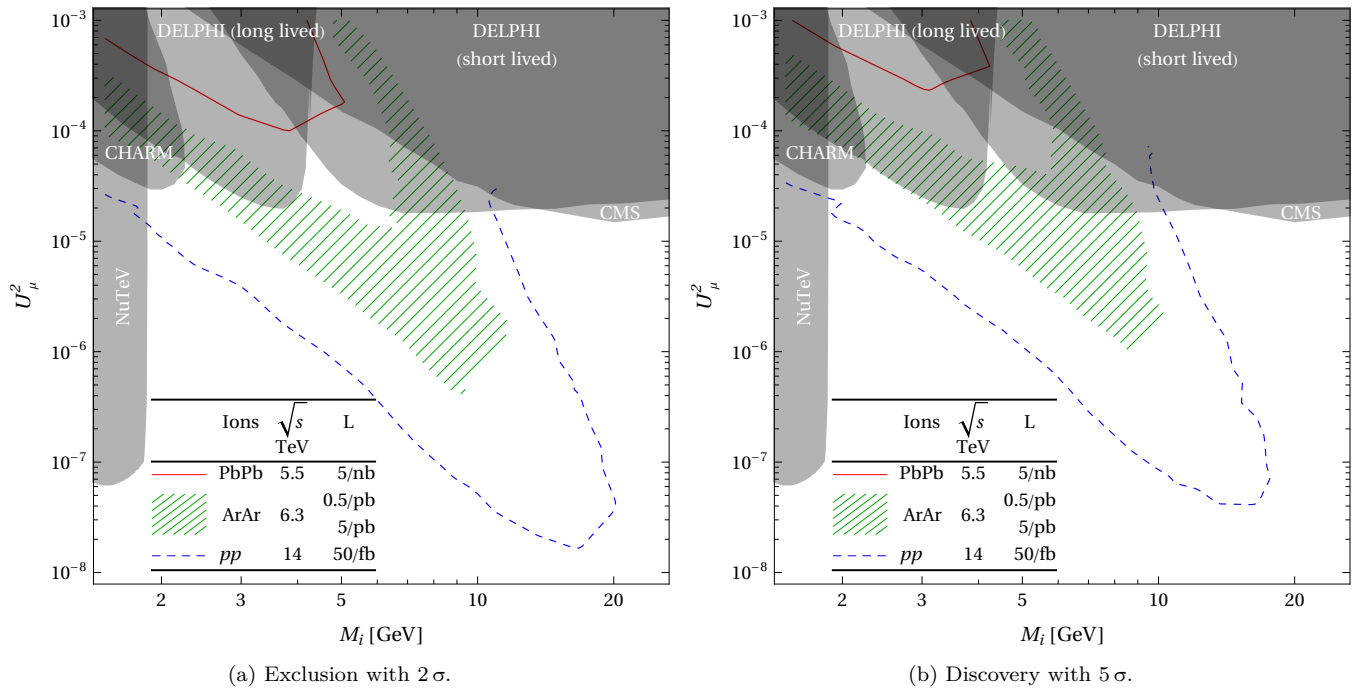

\begin{subfigure}[b]{.495\linewidth}
\graphic{W-boson-exclusion}
\caption{
Exclusion with \unit[2]{\sigma}.
} \label{fig:W boson exclusion}
\end{subfigure}
\hfill
\begin{subfigure}[b]{.495\linewidth}
\graphic{W-boson-discovery}
\caption{
Discovery with \unit[5]{\sigma}.
} \label{fig:W boson discovery}
\end{subfigure}
\caption{
Sensitivity of the CMS detector for heavy neutrinos produced in $W$ decays, as a function of the heavy neutrino mass $M$ and mixing $U_\mu^2$, in PbPb (solid red), ArAr (green hashed band), and $pp$ (dashed blue) collisions at \unit[5.5, 6.3, and 14]{TeV}, respectively.
The luminosities are indicated in the plot and roughly correspond to equal running time of one month.
The left and right panel correspond to exclusion (4 events) and discovery (9 events), respectively.
We expect a comparable result for the ATLAS detector.
The result are based on a simulation of $W$ induced processes using \software{MadGraph5\_aMC@NLO} with the parameters described in \cref{sec:W boson decays}.
The green band reflects the current uncertainty in the beam intensity that can be achieved in ArAr collisions.
The gray areas represent the exclusion limits of former experiments NuTeV \cite{Vaitaitis:1999wq}, CHARM \cite{Bergsma:1985is}, DELPHI \cite{Abreu:1996pa} and CMS \cite{Sirunyan:2018mtv}.
We do not display a constraint on the mixing angle from the requirement to generate the light neutrino masses because light neutrino oscillation data only imposes a lower bound on the mixing of an individual heavy neutrino species if one makes additional model dependent assumptions \cite{Drewes:2019mhg}.
} \label{fig:W boson sensitivity}
\end{figure*}

Following the approach in reference \cite{Drewes:2019fou}, we work under the assumption that the \SM background can be efficiently excluded by the cuts on the invariant mass and the displacement, \cf \cref{fig:background}.
Quantifying the remaining backgrounds would require a very realistic simulation of the whole detector.
These include cosmic rays and beam-halo muons, which only occur at a low rate in the experimental caverns and can mostly be recognized \cite{Liu:2007ad}, as well as scattering of \SM neutrinos from the collision point with the detector, which have a low cross section of charged-current interaction in the detector material.
In summary, we assume that the background number is smaller than one and do a (under this assumption) conservative statistical analysis with one background event, using the non-observation of four events and the observation of nine events for exclusion and discovery, respectively.

\subsubsection{Results}

We present our results in \cref{fig:W boson sensitivity}.
It shows that the suppression of the number of events due to the reduced instantaneous luminosity of heavy ion runs compared to proton runs overcompensates the $A^2$ enhancement per collision, \ie \cref{it:interactions}, so that Pb collisions are clearly not competitive.
For lighter nuclei like Ar the perspectives are somewhat better, as the expected number of events per unit of running time is only about an order of magnitude smaller than in proton runs.
If the heavy neutrinos have mixing angles slightly below the current experimental limits, then they would first be discovered in proton collisions, but heavy ion collisions would still offer a way to probe the interactions of the new particles in a very different environment.
For heavy neutrinos that are produced in $W$ boson decays, the sensitivity is only marginally increased when lowering trigger thresholds, \ie \cref{it:trigger}, because most $\mu^\pm$ from the primary vertex have $p_T > \unit[25]{GeV}$ due to the mass of the $W$ boson.
It remains well below what can be achieved in proton collisions at the LHC \cite{Helo:2013esa, Izaguirre:2015pga, Gago:2015vma, Dib:2015oka, Cottin:2018kmq, Antusch:2017hhu, Cottin:2018nms, Abada:2018sfh, Drewes:2019fou, Bondarenko:2019tss, Liu:2019ayx, Dib:2019ztn, Dib:2018iyr, Cvetic:2018elt, Cvetic:2019rms}.

\subsection{Heavy neutrinos from $B$ meson decays} \label{sec:B decays}

\begin{figure*}
\begin{subfigure}[t]{.49\linewidth}
\graphic{pT}
\caption{
Differential cross section $\dv*{\sigma_B^\text{XX}}{p_T}$ of $B$-mesons.
} \label{fig:B meson pT}
\end{subfigure}
\hfill
\begin{subfigure}[t]{.49\linewidth}
\graphic{muon}
\caption{
Comparison of normalized differential cross sections $\dv*{\sigma^\text{XX}}{p_T \sigma^\text{XX}}$.
} \label{fig:comparison of pT distributions}
\end{subfigure}
\caption{
\emph{Panel \subref{fig:B meson pT}:}
Parton level differential cross section $\dv*{\sigma_B^\text{XX}}{p_T}$ of $B$ mesons per nucleon (including theoretical uncertainties) expected in collisions of the ions indicated in the plot.
\emph{Panel \subref{fig:comparison of pT distributions}:}
Comparison between the differential cross sections of the $B$-mesons and those of the leading muons produced in their decay, each normalized to the respective integrated cross section.
The estimates are based on proton collisions with the collision energy per nucleon $\sqrt s$ in heavy ion collisions \unit[14]{TeV} (dotted blue) for $pp$, \unit[7]{TeV} (dashed green) for ArAr and \unit[5.5]{TeV} (solid red) for PbPb.
As shown in \cref{fig:comparison} the nuclear modification effects in good approximation only lead to an overall rescaling of the cross section, which has little effect on the comparison of event numbers for different $p_T$ cuts described in \cref{sec:Number of events}.
All processes correspond to central collisions with $\abs{\eta} < 4$.
The predictions have been derived with the \software{FONLL} framework \cite{Cacciari:1998it, Cacciari:2001td, Cacciari:2015fta} at $\NLO+\NLL$, using for the $b$-quark fragmentation fraction a value of $f(b \to B^+) = 0.403$ \cite{Cacciari:2012ny} and the \software[6.6]{CTEQ} \NLO parton distribution functions \cite{Nadolsky:2008zw}.
The dotted and dashed vertical lines at \unit[25 and 3]{GeV} correspond to the single muon trigger thresholds in proton and heavy ion collisions, respectively.
} \label{fig:differential cross section}
\end{figure*}

\begin{figure}
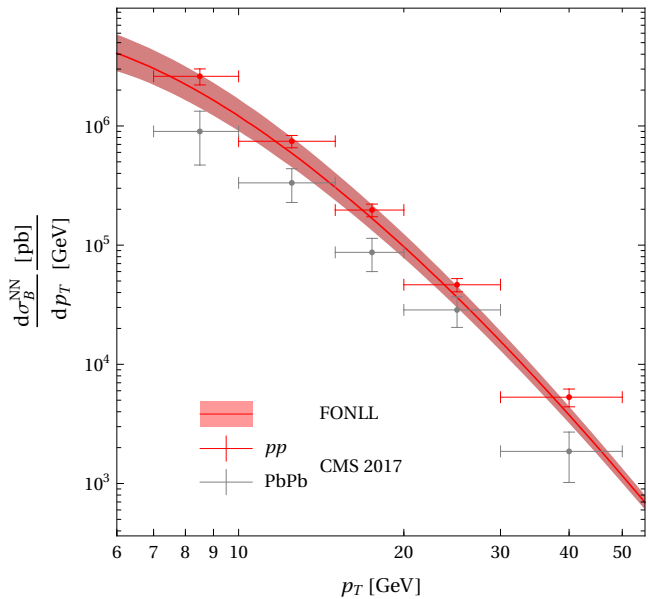

\graphic{comp}
\caption{
Comparison between the theoretical differential production cross sections for $pp \to B^+ + X$ at $\sqrt s = \unit[5.03]{TeV}$ using $\abs{\eta} < 2.5$ computed with \software{FONLL} \cite{Cacciari:1998it, Cacciari:2001td, Cacciari:2015fta} as described in the caption of \cref{fig:differential cross section} (red band) and the experimental measurements for proton and lead collisions at $\sqrt s = \unit[5.02]{TeV}$ (red and gray data points, respectively) \cite{Sirunyan:2017oug}.
The difference of \unit[1]{GeV} between the experimental and the theoretical center-of-mass energy is caused by a technical limitation of \software{FONLL} and the resulting uncertainty is negligible in comparison to the theoretical and experimental errors.
} \label{fig:comparison}
\end{figure}

The situation is very different for heavy neutrinos produced in $B$ meson decays.
The cut-off in sensitivity along the $M$ axis in this case is not determined by the fact that the $N$ decays too quickly to give a displaced vertex signal, but by kinematics: The production cross section exhibits a sharp cut when $M$ approaches the $B$ meson mass $m_B$.
Since this cut occurs in a mass range where the expression \eqref{eq:number W events} suggests that the sensitivity should still improve when increasing $M$, \cf \cref{fig:illustrative plot}, we expect that one can achieve maximal sensitivity just below the threshold.
This means that the sensitivity is maximal in a region where the momenta in the $B$ meson rest frame of both, the $N$ and the $\mu^\pm$ that is produced along with it, are much smaller than $m_B$.
The $p_T$ distribution of $B$ mesons in the laboratory frame peaks around \unit[3]{GeV}, \cf \cref{fig:B meson pT}.
As a result, the vast majority of $\mu^\pm$ have $p_T$ well below standard $p_T$ cuts.
Hence, there is an enormous potential for improving the sensitivity if one can lower the trigger thresholds on the primary muon $p_T$.
For $B$ meson induced processes in heavy ion collisions we assume a trigger threshold of \unit[3]{GeV}, which roughly corresponds to the kinematic limits dictated by the magnetic bending and the geometry of tracking detectors.

The production of heavy neutrinos in $B$ meson decays cannot be simulated in the same way as described in \cref{sec:W boson decays}.
A detailed simulation of $N$ production from $B$ mesons and their decay is technically challenging and goes beyond the scope of this work, the main purpose of which is to estimate the order of magnitude of the sensitivity that can be reached in heavy ion runs.
Therefore, we resort to a modification of the simplified detector model \eqref{eq:number W events} to determine the number of events.
If the masses of all final state particles were negligible, we could express $\sigma_\nu = \sigma_B / 9$, where $\sigma_B$ is the total $B$ meson production cross section and the factor $1/9$ accounts for the branching ratio of the decay into final states including neutrinos.
There is a wide range of \SM two and three body decays into neutrinos, in all of which the \SM neutrino could be replaced by a $N$.
In the two body decay $B^\pm\to \mu^\pm N$ the heavy neutrino mass can be taken into account by multiplying a simple phase space factor,
\begin{equation}
N_\text{obs} =
\frac{L \sigma_B}{9}
\left(1 - \frac{M^2}{m_B^2}\right)^2 U_\mu^2
\left(e^{\nicefrac{-l_0}{\lambda_N}} - e^{\nicefrac{-l_1}{\lambda_N}} \right)
f_\text{cut}
\ .\label{eq:B event number}
\end{equation}
While this decay is helicity suppressed in the \SM, this is not the case for the decay into heavy neutrinos.

\subsubsection{Matching the simplified detector model to simulations}

\begin{figure*}
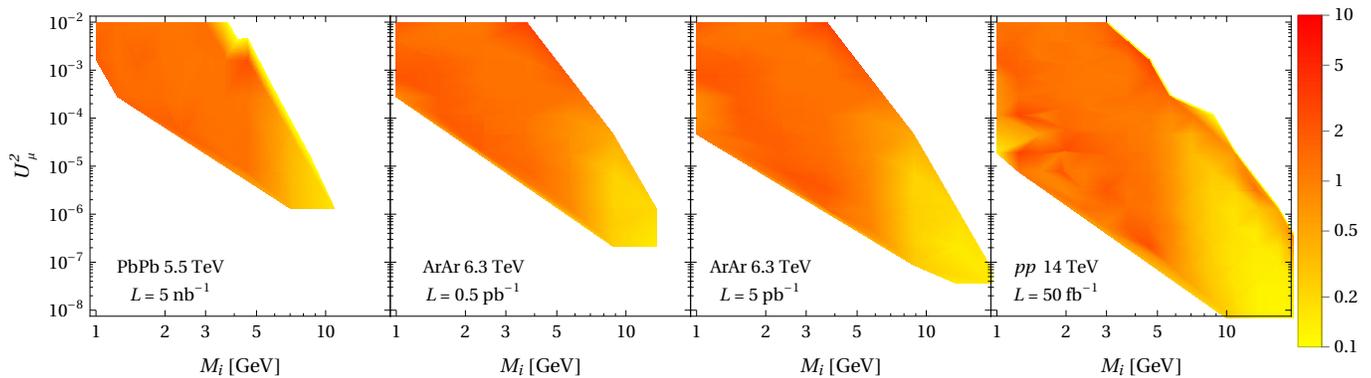

\graphic{Errors}
\caption{
Ratio between the simplified detector model \eqref{eq:number W events} and the number of events predicted by the simulations performed in \cref{sec:W boson decays}, for PbPb, ArAr (pessimistic and optimistic) and $pp$ at \unit[5.5, 6.3, and 14]{TeV}, respectively.
The displayed region in the plane of heavy neutrino mass $M$ and mixing $U_\mu^2$ corresponds to parameter values where the model \eqref{eq:number W events} predicts more than 0.1 events.
} \label{fig:errors}
\end{figure*}

We determine the parameters $l_0$, $l_1$ and $f_\text{cut}$ in the model \eqref{eq:B event number} by fitting the simplified detector model \eqref{eq:number W events} to the results of our simulations for $N$ production in $W$ decays shown in \cref{fig:W boson sensitivity}.
This corresponds to modeling the LHC detectors ATLAS or CMS as spherical, which turns out to be a good estimate up to factors of 2--3, \cf \cref{fig:errors}.
For the neutrino production cross section $\sigma_\nu$ in $W$ boson decays we use the results from \software{MadGraph5\_aMC@NLO}, \ie, $\unit[1.12 \cdot 10^4]{pb}$ for proton collisions at \unit[14]{TeV} and $\unit[40^2 \cdot 4898]{pb}$ and $\unit[208^2 \cdot 4228]{pb}$ for Ar at \unit[6.3]{TeV} and Pb at \unit[5.5]{TeV}, respectively.

In order to account for the Lorentz factor $\beta\gamma$ for each choice of $M$, we compute the $N$ momentum in the laboratory frame as a function of the $W$ boson momentum and the angle between the spacial $W$ and $N$ momenta.
We then average equation \eqref{eq:number W events} over $W$ momenta, using a distribution which we have generated simulating the process $pp \to W$ with up to two jets using \software{MadGraph5\_aMC@NLO} with subsequent hadronization and matching with soft jets via \software{Pythia}.
With $l_0 = \unit[2]{cm}$ and $l_1 = \unit[20]{cm}$ we can reproduce the results of our simulation shown in \cref{fig:W boson sensitivity} in good approximation if we set the overall effective efficiency to $f_\text{cut} = 0.1$.

The fitted parameter values can be understood in terms of physical arguments.
The choice $l_0 = \unit[2]{cm}$ is qualitatively in good agreement with what one would expect from the geometrical cuts $\unit[0.5]{cm}$ and $\unit[10]{cm}$ on the minimal displacement in transversal and longitudinal direction that were used in the simulation.
$l_1 = \unit[20]{cm}$ indicates a typical distance at which one can still reconstruct the displaced vertex.
In the simulation we assumed that the vertex reconstruction efficiency linearly drops from \unit[100]{\%} to zero between a displacement of \unit[5]{mm} and \unit[55]{cm}, hence \unit[20]{cm} is a reasonable value.
The fact that all of the parameter values can be understood physically provides a strong self-consistency check for our approach.
In \cref{fig:errors} we show the ratio between the simplified detector model \eqref{eq:number W events} and the results of the simulation described in \cref{sec:simulation}
within the region where equation \eqref{eq:number W events} predicts more than 0.1 events.
Given the non-linear dependence of the function \eqref{eq:number W events} on the parameters and the fact that $N_\text{obs}$ changes over six orders of magnitude within this region, it is absolutely non-trivial that the simplified model reproduces the simulation up to a factor 2--3 within that region.

\subsubsection{Computing the number of events} \label{sec:Number of events}

\resetacronym{NLO}
\resetacronym{NLL}

In order to determine $\sigma_B$ in the model \eqref{eq:B event number} we first compute the differential cross section $\dv*{\sigma_B}{p_T}$ for $B$ mesons produced at different collision energies at \NLO and \NLL within the \software{FONLL} framework \cite{Cacciari:1998it, Cacciari:2001td, Cacciari:2015fta}, in the range $p_T \in \unit[{[0, 300]}]{GeV}$ using $f(b \to B^+) = 0.403$ for the $b$-quark fragmentation fraction \cite{Cacciari:2012ny} and the \software[6.6]{CTEQ} \NLO parton distribution functions \cite{Nadolsky:2008zw}, accepting events with a pseudorapidity $\abs{\eta} < 4$.
The results are shown in \cref{fig:B meson pT}.
We validate the predictions against experimental results \cite{Sirunyan:2017oug}, noticing that the data are mostly centered on the upper side of the theoretical uncertainty band, \cf \cref{fig:comparison}.
By using central value predictions we are thus underestimating the differential cross section, and the derived results can be interpreted as being conservative.

We fix the value of $\sigma_B$ by integrating over $\dv*{\sigma_B}{p_T} = A^2 \dv*{\sigma_B^\text{XX}}{p_T}$, where the integration limits have to be fixed by the $p_T$ cuts.
We can incorporate the lower $p_T$ cut in heavy ion collisions compared to proton collisions, \cref{it:trigger}, by computing $\sigma_B$ as an integral over $\dv*{\sigma_B}{p_T}$ with different lower integration limits that reflect the different $p_T$ cuts on the primary muon.
The $p_T$ distribution of the primary muons depends on $M$ and should be determined in a simulation.
We take a much simpler approach that gives a very conservative estimate of the discovery potential in heavy ion collisions.
The $B$ meson $p_T$ distribution is a good proxy for the $p_T$ distribution of the leading muon if the heavy neutrino has a mass comparable to the $B$ meson, since in this case the muon will be soft in the $B$ meson rest-frame.
For smaller $M$ the two distributions can differ considerably due to the muon momentum in the $B$ rest frame.
The modification is most extreme for $M = 0$, in which case the kinematics is the same as if the $N$ in the final state is replaced by a \SM neutrino.
One can thus determine the muon $p_T$ distribution in the extreme cases $M = m_B$ and $M = 0$ by computing the $p_T$ distribution of the $B$ mesons themselves as well as that of muons produced in their decay $B^+ \to \mu^+ \nu_\mu\ X$, as shown in \cref{fig:comparison of pT distributions}.
In the second case the distribution peaks at considerably lower $p_T$.

To keep the analysis simple and conservative, we directly apply the experimental $p_T$ cut on the leading muon to the $p_T$ distribution of the $B$ meson when computing $\sigma_B$ from $\dv*{\sigma_B}{p_T}$.
For $\sigma_B$ in proton collisions we use $\unit[25]{GeV} < p_T < \unit[300]{GeV}$, in heavy ions collisions we use $\unit[3]{GeV} < p_T < \unit[300]{GeV}$.
$p_T$ values below \unit[3]{GeV} are very hard to access even in heavy ion collisions because the CMS magnetic field prevents particles with such low momentum from reaching the detector in most of the solid angle range where it is sensitive.
\footnote{
In principle one should consider an $\eta$-dependent $p_T$ threshold.
Realistic numbers for CMS in the same heavy ion environment and a similar muon kinematics can be found in a recent paper based on PbPb data collected in 2015 \cite{Sirunyan:2017oug}, where it is stated that the muon thresholds are $p_T > \unit[3.5]{GeV}$ for $\abs{\eta} < 1.2$, $p_T > \unit[1.8]{GeV}$ for $2.1 < \abs{\eta} < 2.4$, and linearly interpolated in the intermediate $\abs{\eta}$ region.
In order to keep things simple and less specific to the geometry of a specific detector, we use the conservative estimate \unit[3]{GeV}.
}
If we had applied the same cuts to the muon distribution in $B^+ \to \mu^+ \nu_\mu\ X$ decays the predicted number of events in heavy ion collisions would improve by more than an order of magnitude compared to our conservative estimate.
The actual value of $\sigma_B$ lies in between these extreme cases.
All other cuts and efficiencies are summarized in $f_\text{cut}$ and should be similar for proton and heavy ion collisions, except for a sub-dominant change due to the fact that the momentum distributions in heavy ion collisions are slightly different.
Therefore, we adapt the value $f_\text{cut} = 0.1$ obtained from fitting the simplified detector model \eqref{eq:number W events} to the simulation.

We finally take account of the Lorentz factor in the model \eqref{eq:B event number} by expressing $\beta\gamma$ for each choice of $M$ in terms of the $B$ meson momentum in the laboratory frame and the angle between this momentum and the $N$ momentum.
We average equation \eqref{eq:B event number} over both, using a flat prior for the angle in the $B$ rest frame, and adopt $B$ meson spectra that we have determined by generating the process $pp \to \overline b b$ with up to one additional jet using \software{MadGraph5\_aMC@NLO} with subsequent hadronization and matching of soft jets with \software{Pythia}.

\subsubsection{Results}

\begin{figure*}
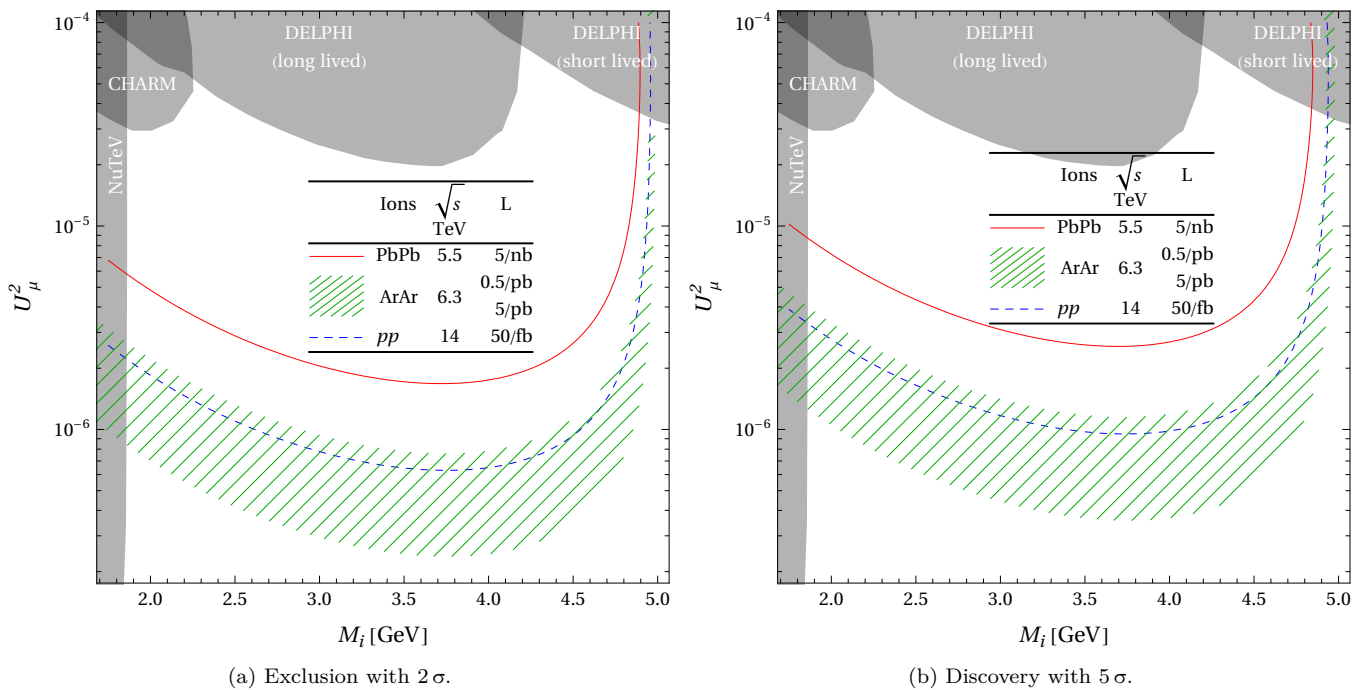

\begin{subfigure}[b]{.495\linewidth}
\graphic{B-meson-exclusion}
\caption{
Exclusion with \unit[2]{\sigma}.
} \label{fig:B meson exclusion}
\end{subfigure}
\hfill
\begin{subfigure}[b]{.495\linewidth}
\graphic{B-meson-discovery}
\caption{
Discovery with \unit[5]{\sigma}.
} \label{fig:B meson discovery}
\end{subfigure}
\caption{
Sensitivity for heavy neutrinos produced in $B$ decays with the same conventions as in \cref{fig:W boson sensitivity}.
The results were obtained from the simplified detector model \eqref{eq:B event number} as described in \cref{sec:B decays}.
} \label{fig:B meson sensitivity}
\end{figure*}

We present the results of our computation in \cref{fig:B meson sensitivity}, where we compare the sensitivity that can be achieved in proton and heavy ion collisions for equal running time using the luminosities as in \cref{sec:W boson decays}.
The results show that data from PbPb collisions could improve existing bounds on the properties of heavy neutrinos by more than an order of magnitude.
Furthermore, for ArAr collisions, the combined enhancement due to the larger number of nucleons, \cref{it:interactions}, and the lower cut in $p_T$, \cref{it:trigger}, can overcompensate the effect of the lower instantaneous luminosity compared to proton collisions, and one can achieve a better sensitivity per unit of running time.
Here we have not taken advantage of the absence of pileup at all, \ie, \cref{it:pileup}, and we recall that we made a very conservative estimate of the effective $\sigma_B$.
This suggests that, in the range $\unit[2]{GeV} < M < \unit[5]{GeV}$ and using the sensitivities estimated by CERN's \emph{Physics Beyond Colliders Working Group} \cite{Beacham:2019nyx}, ArAr collisions with $L = \unit[5]{\inv{pb}}$ could achieve a higher sensitivity than FASER2 with $L = \unit[3]{\inv{ab}}$ and a comparable sensitivity as \mbox{CODEX-b} \cite{Gligorov:2017nwh} with $L = \unit[300]{\inv{fb}}$, while MATHUSLA \cite{Chou:2016lxi, Curtin:2018mvb, Alpigiani:2018fgd} with $L = \unit[3]{\inv{ab}}$ would be more than an order of magnitude more sensitive.
Also the SHiP experiment \cite{Anelli:2015pba,Alekhin:2015byh} with $2 \cdot 10^{20}$ protons on target could achieve a higher sensitivity.
However, no decision has been made so far about the construction of these proposed future detectors.
For FASER the first phase has been approved, which is almost an order of magnitude less sensitive than FASER2 \cite{Ariga:2018uku}.

\section{Discussion}

We propose to search for \LLPs via displaced vertex searches in heavy ion collisions at the LHC.
In the context of \LLP searches heavy ion collisions provide four main advantages in comparison to $pp$ collisions:
\begin{inlinelist}
\item[\ref{it:interactions}] The number of parton level interactions per hadron-hadron collision is larger
\item[\ref{it:pileup}] There is no pileup, which \eg renders the probability of mis-identifying the primary vertex practically negligible
\item[\ref{it:trigger}] The lower instantaneous luminosity makes it possible to considerably loosen the triggers used in the main detectors
\item[\ref{it:fields}] There are new production mechanisms
\end{inlinelist}
The track multiplicity, which is traditionally considered to be a reason that speaks against New Physics searches in heavy ion collisions, is not considerably higher than in high pileup $pp$ collisions, leaving the lower instantaneous luminosity as the main disadvantage.

\subsection{Summary of the main results}

In the present work, we focus on \cref{it:interactions,it:trigger}, using the specific case of heavy neutrinos with masses in the GeV range as an illustrative example.
We consider two production mechanisms of heavy neutrinos, production in $W$ boson decay and in $B$ meson decay.
If the same cuts are applied as in $pp$ collisions we find that the limitations on the instantaneous luminosity for PbPb suppress the observable number of events per unit of run time by almost two orders of magnitude.
This suppression can be reduced to less than one order of magnitude for lighter nuclei, the use of those is currently explored by the heavy ion community for other reasons \cite{Jowett:2018jj} such as the longer beam lifetime.

For the production in $W$ boson decays this means that heavy ion collisions in general do not offer a competitive alternative to searches in proton collisions, though the integrated luminosity of the \HLLHC in ArAr collisions would be sufficient to push the sensitivity far beyond current experimental limits, \cf \cref{fig:W boson sensitivity}.
Loosening the triggers for $N$ produced in $W$ decays only leads to a marginal improvement.
The situation is much more promising when considering the production in $B$ meson decays, which leads to a larger number of events, but signatures with much lower $p_T$.
The results shown in \cref{fig:B meson sensitivity} are remarkable in several ways:

First, data from the complete PbPb run could improve the sensitivity of searches for heavy neutrinos by more than an order of magnitude in comparison to current bounds.
For a small range of masses over \unit[4]{GeV} the improvement would amount to two orders of magnitude.
If the LHC's heavy ion runs were performed with Ar instead, the improvement would be up to three orders of magnitude.

Second, the sensitivity that could be achieved in a given unit of running time is actually larger in ArAr collisions than in proton collisions due to the lower cuts on $p_T$ that can be imposed.
This is not sufficient to entirely compensate for the longer scheduled running time for proton collisions.
However, we did not take advantage of the absence of pileup, \cref{it:pileup}, in the present analysis.
This suggests that for models where pileup poses a serious problem for the extraction of signatures, \cf \eg \cref{fig:misidentification}, heavy ion collisions could actually be more sensitive than proton collisions.

Finally, heavy ion collisions would allow to study the properties of long lived particles in a very different environment than proton collisions.
This can be particularly interesting for cosmologically motivated \LLPs because this environment roughly resembles the primordial plasma that filled the early universe.
The properties of some new particles, such as axion-like particles, are expected to change qualitatively during the transition from \QGP to hadronic matter.
Others, \eg sexaquarks, would be primarily produced during this transition in the early universe, hence the production in heavy ion collisions would resemble the mechanism that generated them cosmologically \cite{Bruce:2018yzs}.
For the specific case of heavy neutrinos in the \nuMSM considered here, a study of the heavy neutrino properties in the \QGP could help to shed light on their potential impact on \DM production, as discussed in \cref{sec:ModelDependent}.

\subsection{Complementary approaches}

Heavy ion collisions are not the only way to search for low $p_T$ events in the LHC main detectors.
Another opportunity is offered by the so called \enquote{$B$ parking} data of CMS, pioneered at the end of Run~2 \cite{CMS-DP-2012-022, CMS-DP-2019-bparking}.
The parking concept consists of storing for later processing (during a long shutdown) a fraction of the data passing mild thresholds.
With the data parked in 2018, CMS is expected to add $10^{10}$ events with low $p_T$ $B$ mesons \cite{Duarte:2018jd} that did not pass the standard trigger paths.
The same order of magnitude is expected to be achievable by CMS at the end of each Run.
\footnote{
As the parked data can only be processed during a long shutdown, parked triggers are expected to be executed only during the last year of each Run.
}
This should be compared with our estimation of the $pp$ yearly dataset in \HLLHC, amounting to around $7 \cdot 10^{10}$ $B$ mesons passing standard triggers.
We are not considering in our study any additional contribution from parking, as there are no firm plans for parking in future \LHC runs, and future storage capacities are difficult to estimate.
We remark, though, that there is no fundamental limitation preventing the same concept to be used also for Heavy Ion runs, which in the context of our proposal may mean enlarging the dataset with further trigger paths, allowing to consider additional signatures (\eg, an electron and a muon, or a muon and a fully reconstructed hadronic final state).
We do not elaborate further, as that would crucially depend on the details of future implementations of the parking concept.

Alternatively, as proposed in \cite{Nachman:2016nes}, the recorded pileup events could be exploited to discover light new physics.
The number of additional useful events can be written as $N_\text{PU} = \ev{\text{PU}} w t r$, where $\ev{\text{PU}}$ is the average pileup, $w$ is the trigger bandwidth, $t$ is the running time, and $r = \flatfrac{\sigma_B^{pp}}{\sigma_\text{tot}^{pp}}$ indicates the ratio of cross sections between $B$ meson production and total inelastic $pp$ cross section.
For $pp$ runs at the \HLLHC we assume $\ev{\text{PU}} \approx 200$ and use for $w$ the stated goal to record \unit[7.5]{kHz} on tape \cite{Andre:2018ioz}.
Based on past experience, we assume a running time of $t \approx 10^7 s$ each year.
Finally, we calculated $\sigma_B^{pp} \approx \unit[1.2 \cdot 10^8]{pb}$ for $B$ mesons produced with $p_T > \unit[3]{GeV}$ (\cf \cref{sec:Number of events}), which divided by $\sigma_\text{tot}^{pp} \approx \unit[80]{mb}$ \cite{Antchev:2017dia} yields $r \approx 1.5 \cdot 10^{-3}$.
Hence, pileup results in $N_\text{PU} \approx 2.25 \cdot 10^{10}$ additional $B$ mesons that could be exploited in the future to enhance the useful statistics of the $pp$ data for our purposes.
While this additional statistics is not insignificant, it is still affected by all limitations of high-luminosity $pp$ runs that are of relevance for our proposal (but not addressed quantitatively in this study) due to the ambiguity to associate the final state to its original production vertex.

Finally, one may wonder whether asymmetric collisions between protons and heavy ions may offer advantages for \LLP searches.
Compared to PbPb collisions, a much larger luminosity and nucleon center of mass energy can be achieved in $p$Pb collisions \cite{Citron:2018lsq}, while maintaining the advantage of being free of pileup.
However, the cross section is also reduced as the multiplicative factor in partonic cross section only scales as $A$ instead of $A^2$.
Therefore, we estimate that a search in $p$Pb collisions would not be more sensitive than one in PbPb collisions.
We have checked that the increase in center of mass energy has only a marginal effect on the processes we have considered.

\subsection{Model dependent remarks} \label{sec:ModelDependent}

We have used the example of heavy neutrinos to illustrate the potential of heavy ion collisions to search for new particles.
We chose this model for two reasons.
First, its phenomenology is very well known, and some of us have studied similar signatures as the ones considered here in proton collisions \cite{Drewes:2019fou}.
This has the advantage that we are in a good position to isolate the specific effects of heavy ion collisions from other uncertainties in the study.
Second, the choice of the model is conservative with respect to the comparison between proton and heavy ion collisions because it only takes advantage of two out of the four benefits \labelcrefrange{it:interactions}{it:fields}.

Since the heavy neutrinos here only acts as an illustrative example, we refrain from going into too much detail about the model itself and its phenomenology.
We only summarize the most relevant information needed to put the sensitivity lines in \cref{fig:B meson sensitivity} into context and refer the interested reader to Reference \cite{Boyarsky:2009ix} as well as the more recent reviews cited in \cref{fn:reviews}.

Light neutrino oscillation data and the baryon asymmetry of the universe can be explained in the entire white part of the plots in \cref{fig:W boson sensitivity,fig:B meson sensitivity} if there are at least three flavors of heavy neutrinos \cite{Abada:2018oly}.
\footnote{
For two heavy neutrinos baryogenesis requires roughly $(\flatfrac{U^2}{10^{-5}})(\flatfrac{M}{\unit{GeV}}) < 1$ \cite{Antusch:2017pkq, Eijima:2018qke, Boiarska:2019jcw} and is only possible for specific flavor mixing patterns \cite{Hernandez:2016kel, Drewes:2016jae, Antusch:2017pkq}.
These restrictions are lifted for three or more right handed neutrino flavors \cite{Abada:2018oly}.
}
Heavy neutrinos in the mass range considered here can indirectly affect the production of sterile neutrino \DM by generating chemical potentials that trigger a resonant enhancement of the \DM production \cite{Shi:1998km}, but current studies suggest that the required magnitude of these potentials \cite{Asaka:2006nq, Ghiglieri:2015jua, Venumadhav:2015pla} can only be generated for mixing angles that are too small to be accessed by the searches we propose \cite{Canetti:2012vf, Canetti:2012kh, Ghiglieri:2019kbw}.

In addition to an improved sensitivity in the GeV mass region, heavy ion collisions can also help to shed light on the role of heavy neutrinos in cosmology because they offer an opportunity to study their properties in a dense plasma that roughly resembles the early universe.
The generation of lepton asymmetries at temperatures below the electroweak scale is highly sensitive to their mass splitting \cite{Shaposhnikov:2008pf,Canetti:2012vf, Canetti:2012kh, Ghiglieri:2019kbw}, which is subject to thermal corrections.
These late time asymmetries would not affect the baryon asymmetry of the universe, but can lead to the aforementioned resonant production of \DM \cite{Shi:1998km, Asaka:2006nq} that manifests itself in observable modifications of the matter power spectrum, \cf \cite{Adhikari:2016bei,Boyarsky:2018tvu} and references therein.
Hence, heavy ion collisions at least in principle can offer a indirect probe of the \DM production mechanism in the \nuMSM, though it should be added that it is not clear whether the in-medium properties of the heavy neutrinos could be probed at a sufficient accuracy to draw any definite conclusions in the foreseeable future.

\section{Conclusion}

In summary, we find that heavy ion collisions provide a promising way to explore regions of the parameter space of hidden sector models that are hard to probe in proton collisions.
We have shown this explicitly for heavy neutrino searches in the \nuMSM.
For this study we only took advantage of the fact that the LHC main detectors can be operated with looser triggers in heavy ion collisions than in high intensity proton collisions, which improves the sensitivity to \LLPs that decay into particles with low $p_T$.
Another advantage of heavy ion collisions is the absence of pileup, which \eg entirely avoids the problem of vertex mis-identification, \ie, eliminates a systematic limitation in \LLP searches with non-trivial event topology.
We postpone a more detailed study of this aspect to future work.
In addition to this, it is well known that heavy ion collisions can offer entirely new production mechanisms that are absent in proton collisions.
In combination, this provides strong motivation to include potential New Physics searches in the discussion of the future of the heavy ion program at CERN \cite{Bruce:2018yzs}.

\begin{acknowledgments}

We thank Olivier Mattelaer for his valuable contribution to the early stages of the work.
We thank Georgios Konstantinos Krintiras, John Jowett, Fabio Maltoni, Emilien Chapon, Jessica Prisciandaro, Mauro Verzetti, Martino Borsato, Elena Graverini, Giacomo Bruno and Matteo Cacciari for very helpful discussions as well as Albert de Roeck and Steven Lowette for pointing us to relevant material.
Some of the figures in this article have been reproduced or adapted from \cite{Drewes:2018xma}.
We appreciated Fosco Loregian's suggestions on how to improve \cref{fig:misidentification}.
This work was partly supported by the F.R.S.-FNRS under the \emph{Excellence of Science} (EOS) project \no{30820817} (be.h).
This project has received funding from the European Union’s Horizon 2020 research and innovation program under the Marie Skłodowska-Curie grant agreement \no{750627}.

\end{acknowledgments}

\raggedright\bibliography{bibliography}

\dummyacronym{BFPP}
\dummyacronym{EMD}
\dummyacronym{NLO}
\dummyacronym{NLL}

\end{document}